\documentclass[aps,prl,twocolumn,preprintnumbers,superscriptaddress,amsmath]{revtex4}
\usepackage{graphicx} 
\usepackage{amsmath,amssymb,mathrsfs}
\usepackage{verbatim}
\usepackage{color}

\begin{document}
\title{Integrated robust universal linear optics based on white-boxed multi-plane light conversion}
\author{Xin-Yu Song}
\affiliation
{Laboratory of Quantum Information, University of Science and Technology of China, Hefei 230026, China.}
\affiliation{CAS Synergetic Innovation Center of Quantum Information $\&$ Quantum Physics, University of Science and Technology of China, Hefei 230026, China.}
\affiliation{Anhui Province Key Laboratory of Quantum Network, University of Science and Technology of China, Hefei 230026, China.}
\author{Lan-Tian Feng\footnote[1]{fenglt@ustc.edu.cn}}
\affiliation
{Laboratory of Quantum Information, University of Science and Technology of China, Hefei 230026, China.}
\affiliation{CAS Synergetic Innovation Center of Quantum Information $\&$ Quantum Physics, University of Science and Technology of China, Hefei 230026, China.}
\affiliation{Anhui Province Key Laboratory of Quantum Network, University of Science and Technology of China, Hefei 230026, China.}
\affiliation{Hefei National Laboratory, University of Science and Technology of China, Hefei, 230088, China.}
\author{Pan Gong}
\author{Yu-Yang Ding}
\affiliation{Hefei Guizhen Chip Technologies Co., Ltd., Hefei 230000, China.}
\author{Guang-Can Guo}
\author{Xi-Feng Ren\footnote[2]{renxf@ustc.edu.cn}}
\affiliation
{Laboratory of Quantum Information, University of Science and Technology of China, Hefei 230026, China.}
\affiliation{CAS Synergetic Innovation Center of Quantum Information $\&$ Quantum Physics, University of Science and Technology of China, Hefei 230026, China.}
\affiliation{Anhui Province Key Laboratory of Quantum Network, University of Science and Technology of China, Hefei 230026, China.}
\affiliation{Hefei National Laboratory, University of Science and Technology of China, Hefei, 230088, China.}


\begin{abstract}
The integrated universal linear optical processor realized via multi-plane light conversion (MPLC) features low hardware complexity and high fabrication robustness, suitable for diverse photonic processing tasks.
However, the lack of validated calibration protocols limits its universality and potential.
Here, we propose a machine-learning-based calibration method and verify it using an eight-mode silicon photonic processor composed of interleaved multiport directional couplers and phase-shifter arrays.
We characterize coupler transfer matrices and current-phase responses of individual phase shifters.
To validate the robustness, we demonstrate its universality with Haar-random unitary matrices at wavelengths of 1500, 1550, and 1580 nm, and under partial phase-shifter deactivation.
With this white-boxed device, we also demonstrate high-dimensional quantum gates and quantum reservoir computing for handwritten digit classification.
Our work advances integrated MPLC-based universal linear optics and paves the way for large-scale, fabrication-robust, broadband hardware for classical and quantum photonic information processing.

\end{abstract}

\maketitle

\section{Introduction}
\label{sec:introduction}
Universal linear optics implements arbitrary unitary transformations across optical modes~\cite{Reck1994Unitary,Clements2016Optimal,Carolan2015Universal} and forms the foundation for photonic quantum computing~\cite{Carolan2015Universal,Arrazola2021,Maring2024,Yu2026}, quantum simulation~\cite{Sparrow2018,Harris2017Transport}, and optical neural networks~\cite{Shen2017DeepLearning}. 
Photonic integration miniaturizes such capabilities onto compact chips with stable phase and high integration density, where electronically tunable elements enable dynamic reconfiguration for diverse classical and quantum photonic tasks within a single processor~\cite{Harris2018Linear,Wang2020IntegratedQuantum,Bogaerts2020Programmable,10.1093/nsr/nwag598}.
To further broaden the applicability of universal linear optics, the hardware must support scaling to a larger number of optical modes while maintaining low loss and broad bandwidth. It should also remain robust against fabrication imperfections, loss imbalance, and even failure of some tunable components.


In contrast to conventional triangular~\cite{Reck1994Unitary} and rectangular~\cite{Clements2016Optimal} mesh networks of Mach–Zehnder interferometers (MZIs), multi-plane light conversion (MPLC) provides such a robust architecture for programmable linear optics by interleaving phase-modulation layers with mode mixing layers~\cite{Morizur2010MPLC,Saygin2020Robust,Tang2017MMI,Tanomura2022Scalable,Markowitz2023AutoCalibrating}.
The MPLC-based architecture mitigates cascaded error accumulation inherent to MZI meshes, while inverse optimization approaches further optimize phase settings to match target unitaries and account for system nonidealities.
To implement universal linear optics, arbitrary target unitary matrices must be rapidly mapped to on-chip phase settings. This mapping requires the prior calibration of the mixing matrices and the electrically actuated response of the individual phase shifters.
However, cascaded mixing makes it difficult to isolate individual component responses using only intensity measurements acquired at the chip output ports.
As a result, the experimental demonstrations rely on device-specific optimization and remain limited to selected operations~\cite{Tanomura2020InP,Tanomura2023DualPolarization,Friedman2025SpaceFrequency,Zelaya2025StateGeneration}.
Although some calibration methods have been proposed, such as model reconstruction~\cite{Bantysh2023Reconstruction,Bantysh2024Intensity}, these approaches impose stringent measurement requirements, and the establishment of an accurate and experimentally accessible mapping between phase settings and optical transformations remains a central practical challenge for integrated MPLC-based processors.
Here we present a machine-learning-based, multistep white-box calibration method and verify it using an eight-mode silicon photonic MPLC processor composed of interleaved multiport directional couplers and phase-shifter arrays.
The method determines, layer by layer, the transfer matrices of all directional couplers and the parameters governing the current-driven phase response of each thermo-optical phase shifter (TOPS).
We calibrate the device independently at wavelengths of 1500, 1550, and 1580 nm, and prove its availability spanning an 80-nm wavelength range with Haar-random
unitary matrices.
At 1550 nm, we further demonstrate that disabling one to seven phase shifters leaves the fidelity of Haar-random transformations nearly unchanged.
Furthermore, we characterize the fidelities of high-dimensional quantum gates and demonstrate quantum reservoir computing for handwritten-digit classification.
Our work advances integrated MPLC-based universal linear optics and paves the way for robust and broadband photonic information processing.

\section{Results}

\subsection{Chip calibration with the machine-learning-based method}

\label{sec:results}
\begin{figure*}[t]
\centering
\includegraphics[width=18cm]{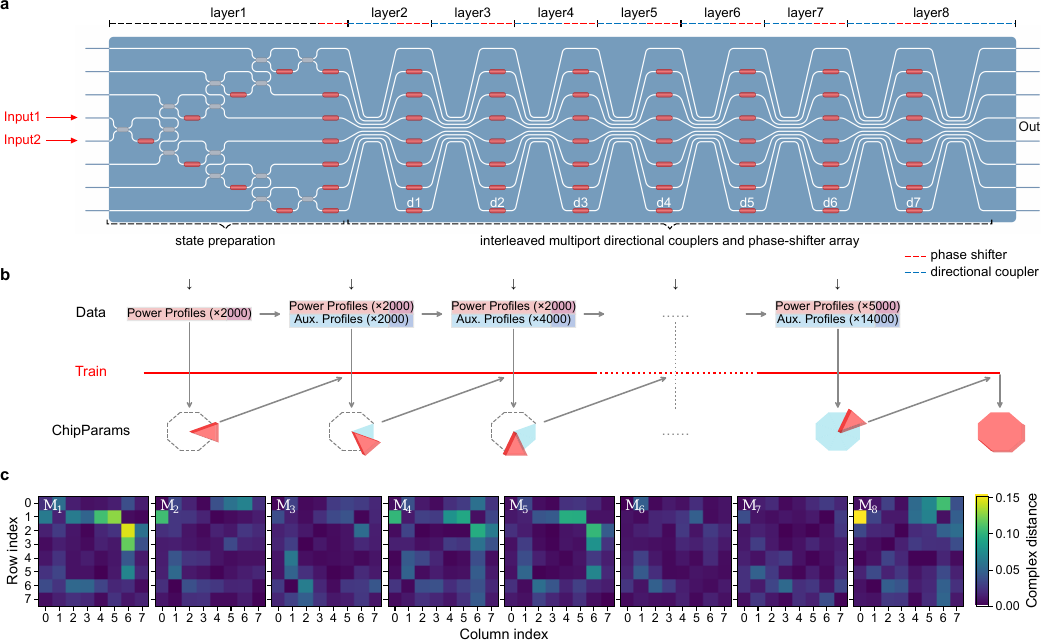}
\caption {Chip structure, calibration method and calibration results.
a, A schematic of the proposed integrated eight-mode MPLC circuit. The red rectangles represent TOPSs, which are individually controlled by a multi-channel direct-current power supply. During the experiments, we employ two input ports and eight output ports. For laser and heralded single-photon characterization, the signal is injected through port 1. In photon-pair measurements, the two photons are injected into ports 1 and 2, respectively.
b, A flow diagram of the algorithm, showing data acquisition, model training, parameter transfer between successive calibration stages and clamping of previously calibrated parameters. Red regions mark tunable parameters optimized during training, whereas blue regions indicate pre-calibrated fixed parameters. c, 
The heatmaps show the element-wise deviations of the eight matrices from their mean. 
}
\label{fig:architecture}
\end{figure*}

Figure~\ref{fig:architecture}a shows the architecture of the proposed integrated MPLC circuit with eight spatial modes. 
The input stage consists of seven reconfigurable MZIs for state preparation, where each MZI comprises two multimode interferometer beam splitters and one TOPS.
The subsequent structure consists of interleaved TOPS arrays and multiport directional couplers.
The multiport directional couplers are designed to be identical for spatial-mode mixing, and the TOPS arrays are programmed to implement arbitrary unitary transformations.
To calibrate the chip, we construct a physics-based forward model that relates the PS responses to the overall transformation: 
\begin{equation}
\label{eq:overall-transformation}
U=P_8M_8P_7M_7\cdots P_1M_1S_{\mathrm{prep}},
\end{equation}
where \(S_{\mathrm{prep}}\) is the transfer matrix of the state-preparation network, \(P_m\)
(\(m=1,\ldots,8\)) are diagonal phase matrices, and
\(M_m\) are the transfer matrices of the directional couplers. Each matrix contains several independent parameters to be determined.

We divide the circuit into eight layers and use a nine-step calibration procedure to determine these parameters (Fig.~\ref{fig:architecture}b). 
In step \(i\) (\(i=1,\ldots,7\)), we generate 2,000 random current configurations, assigning random currents to the PSs in the first \(i\) layers while keeping all remaining PSs at zero current. 
%
With a continuous-wave laser injected into the chip, we apply each current configuration to the chip and record the output power profile across all output ports. 
In the eighth step, we generate 5,000 random current configurations for all PSs and acquire 5,000 samples using the same measurement procedure.
At each step, the newly acquired data form the primary data set, while all data collected in the preceding steps are combined into an auxiliary data set. Both datasets are used for training and testing, with a train-test split ratio of 8:2.
Note that data collection is performed only for the first eight steps. The acquired data are used for machine-learning training to progressively characterize the photonic circuit under the unitary constraint on the directional couplers. 
In the final step, we relax this constraint and refine the model at a lower learning rate to account for mode-dependent loss.



To determine the chip parameters in the actuated layers, we minimize a combined loss function computed from the normalized predicted and measured power profiles:
\begin{equation}
\label{eq:training-loss}
\mathcal L
=
w_1\mathcal L_1
+
w_{\mathrm{dist}}\mathcal L_{\mathrm{dist}}
+
w_{\mathrm{MSE}}\mathcal L_{\mathrm{MSE}},
\end{equation}
where \(\mathcal L_1\) and \(\mathcal L_{\mathrm{MSE}}\) are the mean absolute error and mean squared error, respectively, each scaled by a factor of four, and \(\mathcal L_{\mathrm{dist}}\) is the mean Euclidean distance between the predicted and measured profiles. The weights \(w_1\), \(w_{\mathrm{dist}}\), and \(w_{\mathrm{MSE}}\) are dynamically adjusted based on the current MSE loss, shifting emphasis from \(\mathcal L_{\mathrm{MSE}}\) towards \(\mathcal L_1\) and \(\mathcal L_{\mathrm{dist}}\) as the MSE loss decreases.
We minimize the loss function using the AdamW optimizer, progressively releasing groups of parameters according to their initial scales (see Methods). 
%

Following the above calibration procedure, we extract parameters for all components across the chip, including initial phase offsets of PSs, transfer matrices of multiport directional couplers, output coupling efficiency, and position-dependent loss, among others. In short, the entire circuit is fully white-boxed. As an example, the deviations in the directional-coupler transfer matrices at the wavelength of 1550 nm are presented in Fig. 1c, although the couplers are designed to be identical. With the calibrated chip parameters, we can rapidly map arbitrary target unitary transformations to the driving currents of each PS (see Supplementary Materials).



\subsection{Demonstration of universality under broadband operation and partial phase-shifter deactivation}
\label{sec:robustness}

\begin{figure*}[t]
\centering
\includegraphics[width=18cm]{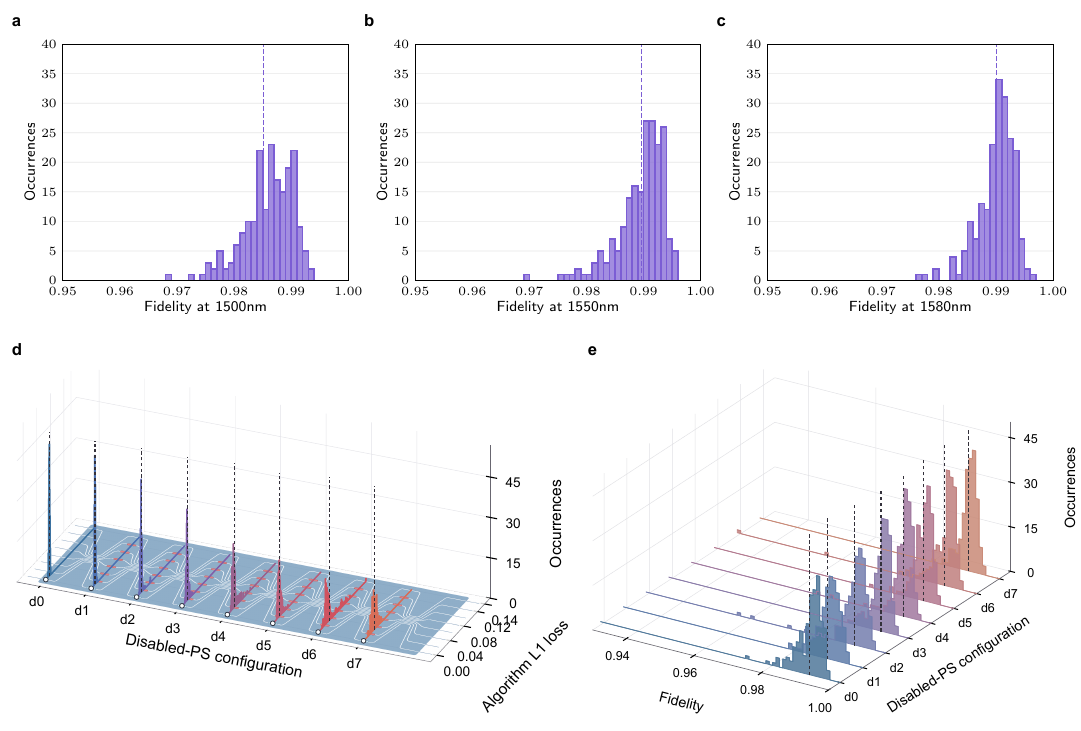}
\caption{Robustness of Haar-random transformation programming across different wavelengths and disabled-PS configurations. a–c, Histograms of the measured fidelity for the same 200 Haar-random transformations at wavelength of 1500, 1550 and 1580 nm, respectively. 
d, Distributions of the residual \(L_1\) loss evaluated from the forward model at the re-optimized driving currents for different disabled-PS configurations. The \(L_1\) loss is defined as
$\mathcal{L}_{1} = \frac{1}{2M}\sum_{i,j=1}^{M}
\left|p^{\mathrm{model}}_{j|i} - p^{\mathrm{target}}_{j|i}\right|$,
where $p^{\mathrm{model}}_{j|i}$ and $p^{\mathrm{target}}_{j|i}$ are the normalized model-predicted and target output probabilities for input $i$, respectively, and $M=8$ is the number of modes. 
e, Distributions of the measured fidelity for different disabled-PS configurations.
}

\label{fig:robustness}
\end{figure*}

To prove the availability of the device with broad bandwidth, we verify its universality at wavelengths of 1500, 1550, and 1580~nm. 
The calibration procedure is carried out separately at different wavelengths, yielding three wavelength-specific sets of chip parameters. The same set of 200 Haar-random unitary matrices is used at all three wavelengths to evaluate the performance of the chip in implementing target transformations with the calibrated models.
With a continuous-wave laser injected into the chip, we program the MZIs in the state-preparation region to sequentially generate all eight spatial input modes. Then we measure the corresponding output powers, obtaining an \(8\times8\) power transmission matrix for each target transformation.
The amplitude fidelity for the target transformation \(n\) is evaluated as
\begin{equation}
\label{eq:amplitude-fidelity}
F_n
=
\frac{1}{8}
\sum_{j=1}^{8}\sum_{i=1}^{8}
\sqrt{
q_{ij}^{(n)}
\left|
\left(U_{\mathrm{target}}^{(n)}\right)_{ij}
\right|^2
},
\end{equation}
where \(q_{ij}^{(n)}\) denotes the probability of
the energy distribution at output port \(i\) for input port \(j\).
After compensation for output coupling efficiency, the mean fidelities are
\(98.52\% \pm 0.94\%\), \(98.96\% \pm 0.41\%\), and
\(99.00\% \pm 0.32\%\) at 1500, 1550, and 1580~nm, respectively,
with the uncertainties representing the standard deviations across the
200 target Haar-random unitary matrices.
These results prove the robustness of on-chip MPLC to wavelength-dependent variations. Not that the device can operate over a much broader wavelength range, and the demonstrated 80 nm range is primarily limited by coupling efficiency (see Supplementary Materials).

Next, we demonstrate universality under partial phase-shifter deactivation with the laser wavelength at 1550 nm. The remaining PS settings can be re-optimized to compensate for the failure of some PSs \cite{Markowitz2023AutoCalibrating}. Within each PS array, we select one PS to mimic device failure (marked in Fig. 1a). This yields eight cumulative disabled-PS configurations, \(d_0,\ldots,d_7\): for each subsequent configuration, one extra PS is disabled, with disabled PSs chosen sequentially across layers from front to back, while all previously disabled PSs remain deactivated, such that \(d_0\) has no disabled PSs and \(d_7\) has seven disabled PSs. 
With the currents of disabled heaters constrained to zero, we re-optimize the driving currents needed for the remaining PSs for each Haar-random target.
The residuals \(L_1\) losses evaluated from the forward model at the optimized currents, averaged over 200 targets, are 0.010346, 0.011277, 0.013703, 0.015610, 0.020571, 0.019796, 0.032191, 0.031742 for \(d_0\) through \(d_7\), respectively (Fig.~\ref{fig:robustness}d). 
We then apply the obtained current settings to the chip and measure the output-power matrix for each target. 
As shown in Fig. 2e, the mean fidelity remains at \(99.10\%\pm0.24\%\) even with seven disabled PSs, showing no significant degradation.

\subsection{Demonstration of quantum gate and reservoir computing}
\label{sec:quantum}
\begin{figure*}[t]
\centering
\includegraphics[width=18cm]{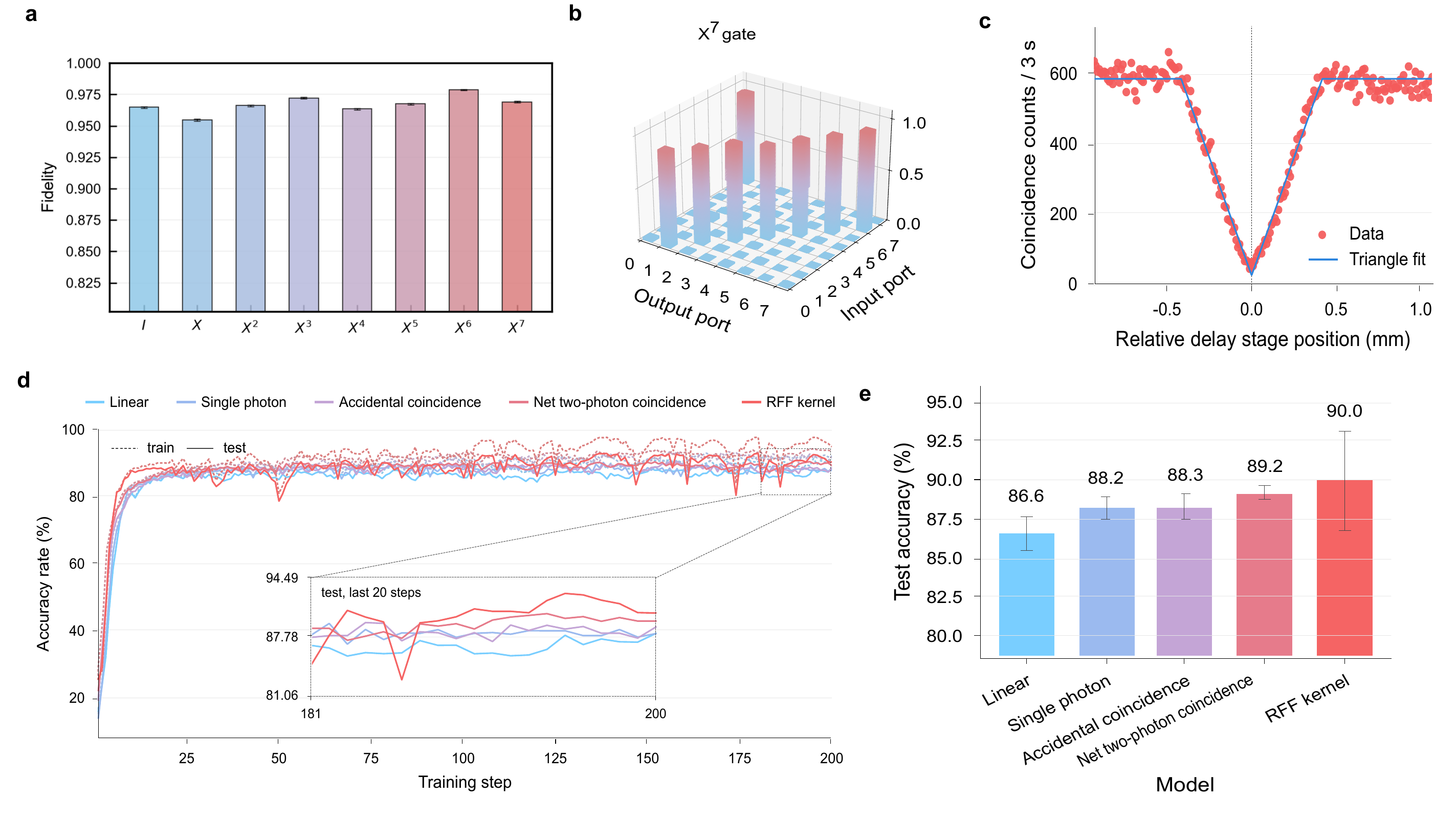}
\caption {Experimental demonstration of the high-dimensional quantum X gate and quantum reservoir computing on the MPLC chip.
a, Fidelities of eight high-dimensional \(X^i\) gates (\(i=0,\ldots,7\)).
b, Output probability distribution of the \(X^{7}\) gate. 
c, HOM interference measured with the chip configured to act as a 50:50 beam splitter. Accidental coincidences have been subtracted. The blue line is a triangle fit to the coincidence counts. 
d, Training and test accuracies as a function of training epochs for the linear baseline, single-photon features, accidental coincidence features, net two-photon coincidence features and random Fourier features (RFF). The inset shows the test accuracies over the last 20 epochs. e, Final test accuracies for the five configurations. For each training run, the test accuracy is averaged over the last 20 epochs. Bar heights and error bars represent the mean and standard deviation, respectively, across ten training runs with different initializations using the same dataset.
}
\label{fig:quantum}
\end{figure*}

Then, we program the MPLC circuit to implement the eight-dimensional cyclic-shift quantum gates \(X^i\) (\(i=0,\ldots,7\)). Their matrix elements are defined by \((X^i)_{rc}=1\) for \(c=(r+i)\bmod 8\) and zero otherwise, where \(r,c=0,\ldots,7\). 
Heralded single photons are injected into the chip for gate characterization (see Methods).
The fidelity of each gate is shown in Fig.~\ref{fig:quantum}a, with an average fidelity of 0.9671. 
Figure~\ref{fig:quantum}b shows the measured \(X^{7}\), and the fidelity is \(0.9690\pm0.0007\), where the uncertainty represents one standard deviation obtained by propagating photon-counting uncertainties.

Finally, we use the MPLC circuit to implement the quantum reservoir computing proposed recently in~\cite{Sakurai2025Reservoir}, and demonstrate it using heralded single photons and indistinguishable photon pairs. 
We characterize the temporal overlap of the two photons using the Hong–Ou–Mandel (HOM) measurement configuration~\cite{PhysRevLett.59.2044}. The relative delay between the two photons is scanned and the measured HOM dip is shown in Fig.~\ref{fig:quantum}c. After subtracting accidental coincidences, the visibility extracted from the triangle fit is \(96.23\pm0.39\%\), with the uncertainty derived from photon-counting fluctuations. The delay position corresponding to the minimum of the measured HOM curve is used for subsequent reservoir computing experiments.
We extract eight principal components from each MNIST image~\cite{Deng2012MNIST} and map them to seven relative phase shifts in a single MPLC PS layer. In the experiment, the PSs used are located in Layer 4 in Fig. 1a. The corresponding driving currents are calculated using the calibrated phase-current relations. The preceding and following sections of the MPLC circuit served as the pre-circuit and the reservoir, respectively, with no currents applied to either section. Coincidence counts are recorded for each current configuration. More details about the preprocessing and encoding procedures are given in Methods.


The measured reservoir features are combined with the original image vectors to train a linear classifier for handwritten-digit recognition.
The test accuracies of the linear baseline, the single-photon model, accidental coincidence reference, net two-photon coincidence model and the RFF model are \(86.65\pm1.08\%\), \(88.17\pm0.73\%\), \(88.29\pm0.79\%\), \(89.17\pm0.47\%\), and \(89.97\pm3.22\%\), respectively. Reported values are means and standard deviations across ten training runs. The observed trends in mean test accuracy are broadly consistent with
the theoretical predictions~\cite{Sakurai2025Reservoir}.

\section{Discussion}
\label{sec:discussion}

To address the calibration challenge of integrated MPLC-based universal linear optical processors, we propose a layer-by-layer, machine-learning-based calibration method. We successfully implemented this method on an eight-mode silicon photonic MPLC processor and demonstrated its effectiveness via a set of applications. Notably, our method holds potential for photonic integrated circuits with complex parameter spaces, which is applicable not only to similar MPLC circuits but also for rapid calibration of large-scale mesh networks.
We note that a recent similar study adopted machine-learning-based global optimization to calibrate a six-mode MPLC photonic processor~\cite{Kondratyev2026Universal}. Compared with the global optimization strategy, our layer-by-layer optimization achieves faster convergence and better robustness against fabrication errors.

The analysis of the calibration data shows that the \(L_1\) loss between the measured and model-predicted output probability distributions exhibits a strong positive correlation with the total electrical power consumption of the chip.
Heating induced by driving currents can modify the transfer matrices of multiport directional couplers, perturbing the output power distributions. Consequently, finer chip temperature control and thermal isolation of individual heaters to suppress thermal crosstalk may further improve the fidelity of on-chip integrated MPLC processors. More research is required to quantify the contributions of these thermal effects and assess the efficacy of such mitigation measures.



In conclusion, we present a white-box calibration method for on-chip MPLC systems. With this approach, we enable reliable operation of the on-chip MPLC processor across broad bandwidths and under partial phase-shifter deactivation. Furthermore, we experimentally validate the practical functionality of the calibrated chip through representative quantum tasks, including high-dimensional quantum gates and photonic quantum reservoir computing. This calibration method possesses broad applicability, and our work demonstrates the high robustness of integrated MPLC-based universal linear optics, paving the way toward large-scale, fabrication-robust, broadband hardware for classical and quantum photonic information processing.


\section{Methods}
\label{sec:methods}


\subsection{Chip fabrication.}
The developed silicon photonic circuit is fabricated using CMOS-compatible UV-lithography processes in a commercial multi-project wafer run. The waveguides are designed to be single moded, having a cross section of 500$\times$220 nm$^2$ with a silicon dioxide upper cladding. TOPSs are used to reconfigure the integrated circuits and each exhibits a resistance of approximately 600 $\Omega$.

\subsection{Machine-learning based calibration method.}

To mathematically describe the overall on-chip transformation, we establish different mathematical models for different parts.

For the on-chip TOPSs, we model the induced relative phase shifts as
\begin{equation}
\label{eq:phase-current}
\boldsymbol{\phi}
=
a\bigl(\mathbf I^{\odot 2}\bigr)C+\mathbf b,
\end{equation}
where \(a\) describes the current-phase coefficient, \(\mathbf b\) describes the phase offset, and \(C\) describes the thermal crosstalk from the PSs to neighboring waveguides \cite{Ceccarelli2020LowPower}.

We assume that each \(2\times2\) multimode interferometer (MMI) has a splitting ratio of 50:50 and a field-amplitude factor \(\alpha_i\) (\(i=1,\ldots,14\)). The transfer matrix of each MMI is expressed as
\begin{equation}
\label{eq:mmi-matrix}
U_{\mathrm{MMI},i}
=
\begin{pmatrix}
\alpha_i & i\alpha_i \\
i\alpha_i & \alpha_i
\end{pmatrix}.
\end{equation}

In part of interlacing PS arrays and directional couplers, each phase shifter layer contains 7 tunable phase shifters and is represented by
\begin{equation}
\label{eq:phase-layer}
P_m
=
\operatorname{diag}
\left(1,e^{i\theta_1^m},\ldots,e^{i\theta_7^m}\right),
\qquad m=1,\ldots,8.
\end{equation}

The transfer matrix of each directional coupler is parameterized as \(U_m=\mathscr U(\boldsymbol{\xi}_m)\) in the unitary case, using a decomposition into elementary unitary transformations \cite{Reck1994Unitary}, and as \(M_m=u_m\odot\exp(iv_m)\) in the non-unitary case, where \(u_m,v_m\in\mathbb R^{8\times8}\), \(|(u_m)_{jk}|<1\), and both the multiplication and exponential are applied element-wise.

As we apply nonzero currents to only \(7(i+1)\) phase shifters (\(i\) represents the calibration step), the unactuated part is represented by \(B(u,v)\in\mathbb C^{8\times8}\), an unconstrained complex matrix. Here, \(u\) and \(v\) are real-valued matrices that parameterize its element-wise amplitude factors and phases, respectively. At step \(i\), the forward function describing the on-chip transformation matrix take the form of
\begin{equation}
\label{eq:stage-transformation}
U
=
B(u,v)P_iU_{i-1}P_{i-1}
\cdots U_1P_1S_{\mathrm{prep}}.
\end{equation}

For the input state \(|4\rangle\) used in our experiment, the parameterized forward model predicts the output power profile as
\begin{equation}
\label{eq:predicted-output}
\mathbf P_{\mathrm{pre}}
=
\operatorname{diag}\!\left[
U(\mathbf I;\boldsymbol{\eta},B)
|4\rangle\langle4|
U^\dagger(\mathbf I;\boldsymbol{\eta},B)
\right],
\end{equation}
where \(\operatorname{diag}[\cdot]\) extracts the diagonal elements as a vector. Since the measured output powers are relative values, both the predicted and measured power profiles are normalized to unit sum before evaluating the training loss.

The loss function in equation~\eqref{eq:training-loss} is then minimized using the AdamW optimizer.
Although we adopt a progressive actuation strategy, we also update the parameters in stages to prevent training stagnation at the outset.
Initially, we clamp the parameters obtained in the previous step.
At each epoch, we allow a subset of these parameters to vary freely by removing the clamping constraints according to the current \(\mathcal{L}_{\mathrm{dist}}\).
Parameters with relatively large initial scales, including the phase offsets \(b\), the amplitude and phase parameters \(u\) and \(v\) of the unactuated part, the directional-coupler parameters \(\boldsymbol{\xi}\), and output coupling efficiencies, are released first, followed by \(\alpha\), \(a\) and \(c\), which have smaller initial scales.

The distance loss remained relatively high after the above-mentioned training procedure, with a normalized value of approximately 0.0616 at 1550~nm.
This residual error may be partly attributable to differences in loss among the eight spatial modes, arising from the different waveguide bend radii and fabrication imperfections.
In the ninth step, we used all existing power profiles and replaced the unitary matrices \(U_m(\boldsymbol{\xi}_m)\) with the non-unitary matrices \(M_m\) to account for the non-unitary response.
Following this training step, the normalized distance loss decreased to 0.031.

\subsection{Quantum reservoir computing with boson sampling.}

To benchmark the model on a real MPLC chip, we randomly selected 7,000 handwritten digit images from the MNIST dataset \cite{Deng2012MNIST}, using 6,000 images as the training set and 1,000 images as the test set. We first scaled the pixel values to \([0,1]\), then applied principal component analysis to retain the first eight principal components. Each component was independently rescaled to \([0,\pi]\) using min–max normalization, yielding the encoding phases \((\theta_0,\ldots,\theta_7)\). We selected one phase-shifter layer of the MPLC circuit for data encoding. Taking \(\theta_0\) as the reference phase, we applied the relative phase shifts \((0,\theta_1-\theta_0,\ldots,\theta_7-\theta_0)\), using their equivalent values within one \(2\pi\) period. This removes an overall phase factor and preserves the output detection probabilities.

For each sample, the reservoir features were normalized to unit sum and then standardized feature-wise using the means and standard deviations calculated from the training set. The same standardization parameters were applied to the test set. The original image vector \(\mathbf{x}\) was then concatenated with the standardized reservoir feature vector \(\mathbf{u}(\mathbf{x})\), yielding the linear classifier input \(\mathbf{y}=(\mathbf{x},\mathbf{u}(\mathbf{x}))\). For the random-coincidence reference, the features were constructed from pairwise products of the singles counts recorded in the indistinguishable two-photon experiment. The linear classifier was trained for 200 epochs using the AdaGrad algorithm with an initial learning rate of \(\eta=0.04\). Training was repeated 10 times with different random initializations and a fixed training–test split. For each run, the accuracy was averaged over the final 20 epochs. The reported accuracies and error bars represent the mean and standard deviation, respectively, of these averages across the 10 runs.

\subsection{Details about experimental test.}

The heralded single photons were produced through type-II spontaneous parametric down-conversion. One photon from each correlated photon pair was sent to an SNSPD as the heralding photon, and the other was injected into spatial mode \(|4\rangle\) and routed using the binary-tree state-preparation circuit.
For gate characterization, the signal photon was routed into each of the eight spatial modes of the MPLC part.

For the single-photon reservoir computing experiments, with the idler photon serving as the heralding photon, the signal photon was routed by the binary-tree state-preparation circuit into spatial mode \(|4\rangle\) at the input of the MPLC part. The single-photon measurements were performed by recording coincidence counts between the heralding detector and the output detectors.

For the two-photon experiments, the signal and idler photons were injected into input ports 4 and 5 of the chip, respectively, and routed by the binary-tree state-preparation circuit to spatial modes \(|4\rangle\) and \(|5\rangle\) at the input of the MPLC part.

To determine the relative delay at which the two photons overlapped temporally, we performed a HOM measurement using the same input ports and routing paths. For this measurement, we configured the first MZI in the binary-tree state-preparation circuit as a balanced 50:50 beam splitter and set the subsequent MPLC part to the identity transformation by applying the corresponding currents.

\section*{Data availability}
Data used in this work are available from the corresponding authors upon reasonable request.


\bibliographystyle{unsrtnat}
\bibliography{references}



\section*{Acknowledgements}
This work was supported by Quantum Science and Technology-National Science and Technology Major Project (Grants 2025ZD0300300, 2021ZD0303200, 2021ZD0301500, 2024ZD0301400), National Key Research and Development Program of China (2022YFA1204704), the National Natural Science Foundation of China (62275240, 62435009, U25D9003, T2325022, U23A2074), the CAS Project for Young Scientists in Basic Research (No. YSBR-049), and the Fundamental Research Funds for the Central Universities. This work was partially carried out at the USTC Center for Micro and Nanoscale Research and Fabrication.

\section*{Author contributions}
XFR and LTF supervised and managed the project. XYS developed the calibration method and carried out the experiments. PG and YYD carried out the packaging and testing of the chip. XYS, GCG, XFR and LTF analyzed the data. All authors contributed to the writing of the manuscript.

\section*{Competing interests}
The authors declare no conflict of interest.

\end{document}